\documentclass[%
 reprint,
 amsmath,amssymb,
 aps
]{revtex4-2}
\usepackage{graphicx}% Include figure files
\usepackage{dcolumn}% Align table columns on decimal point
\usepackage{bm}% bold math
\usepackage{mathtools}
\usepackage{dsfont}

\usepackage[colorlinks,citecolor=red,urlcolor=blue,bookmarks=false,hypertexnames=true]{hyperref} 
\usepackage{siunitx}
\usepackage[version=4,arrows=pgf-filled,mathfontname=mathsf]{mhchem}
\usepackage[gen]{eurosym}
\usepackage{multirow}
\usepackage{float}
\usepackage[stable]{footmisc}
\usepackage{xspace}
\usepackage{multirow}
\usepackage{braket}

\newcommand{\qe}{\ensuremath{Q_\mathrm{e}}\xspace}
\newcommand{\me}{\ensuremath{m_\mathrm{e}}\xspace}

\newcommand{\we}{\ensuremath{\omega_{\mathrm{e}}}\xspace}

\newcommand{\kB}{\ensuremath{k_\mathrm{e}}\xspace}

\begin{document}

\preprint{APS/123-QED}

\title{A 3D-Printed GHz Microwave-Resonator Paul Trap for Electron Confinement and Millisecond Spin-Qubit Coherence}

\author{Niklas V. Lausti$^{1}$}
\thanks{These authors contributed equally to this work.}

\author{Vineet Kumar$^{1,2,3}$}
\thanks{These authors contributed equally to this work.}

\author{Ivan Hud\'{a}k$^{1,4}$}%y

\author{Ji\v{r}\'{i} Hajny\v{s}$^5$}%

\author{Peter K\'{u}\v{s}$^1$}%

\author{Radek Pla\v{s}il$^1$}%

\author{Michal Hejduk$^1$}%
 \email{michal.hejduk@matfyz.cuni.cz}
\affiliation{%
$^1$Charles University, Faculty of Mathematics and Physics, Dept. of Surface and Plasma Science, Prague 8, Czech Republic}%

\affiliation{%
$^2$Center for Quantum Information and Quantum Biology (QIQB), Osaka University, 1-2 Machikaneyama, Toyonaka 560-0043,
Japan}%

\affiliation{%
$^3$Center for Quantum Engineering and Science, QuuQ, Patna 800023, Bihar, India}%

\affiliation{%
$^4$Institute of Photonics and Electronics CAS, v.v.i., Chaberská 1014/57, Prague 8, Czech Republic}%

\affiliation{%
$^5$Faculty of Mechanical Engineering, VŠB - Technical University of Ostrava, Ostrava, Czech Republic}%
 
\begin{abstract}
We report a 3D-printed microwave-resonator Paul trap that operates at the GHz frequencies required for electron confinement while preserving an open geometry suitable for laser access and imaging ion Coulomb crystals. The resonator exhibits moderate quality factors of approximately 1000, enabling large confining electric-field amplitudes at low input power, and is in good agreement with finite-element-method (FEM) predictions despite manufacturing imperfections, which we experimentally characterize. The validated FEM model is then used to calculate the electromagnetic fields in the trapping region. These field distributions, together with surface-roughness measurements, are incorporated into a model of electron spin-qubit decoherence to estimate the coherence time. Our analysis predicts that electron spin-qubit coherence times on the order of \qty{10}{ms} should be achievable under the assumption that motional heating is dominated by Johnson noise, making such systems relevant for quantum information processing. The method of obtaining the coherence time is applicable to any electron Paul trap.
\end{abstract}

\maketitle

%\tableofcontents

\section{Introduction}
\label{sec:intro}

Radio-frequency (RF) Paul traps provide a versatile platform for confining charged particles using purely electric fields, independent of the sign of their charge. This property enables, in principle, the simultaneous trapping of oppositely charged species within a common volume. When the charge-to-mass ratios of the species differ significantly, multiple driving frequencies are required to achieve stable confinement \cite{foot2018two,dehmeltEconomicSynthesisPrecision1995,leefer2017investigation}. Such dual-frequency traps are particularly attractive because they retain the advantages of purely electric trapping schemes, including rapid tunability, scalability, and compatibility with optical spectroscopy due to the absence of strong magnetic fields.

The ability to simultaneously confine light charged particles such as electrons or positrons together with heavy atomic ions or antiprotons opens a range of scientific opportunities. In antimatter research, dual-frequency trapping offers a route to controlled recombination processes and improved production of neutral antimatter. The quantization of the electron motional spectrum within the trapping potential enables state-selective laser-assisted recombination into atomic bound states, potentially increasing the production of desired states while suppressing pathways leading to rapid autoionization. More broadly, hybrid systems combining electrons and ions could provide access to many-body regimes that are inaccessible to either species alone.

Such systems are also of interest for quantum information science. Electrons possess strong coupling to microwave fields and can exhibit fast dynamics owing to their small mass and high motional frequencies, whereas trapped ions provide long coherence times together with high-fidelity optical preparation, manipulation, and readout. Combining both species therefore promises hybrid architectures that exploit the strengths of each platform. In addition, the electron spin can serve as a qubit in a regime detached from an ion, complementing approaches based on condensed-matter devices \cite{kawakamiBlueprintQuantumComputing2023,zhouSingleElectronsSolid2022} and semiconductor spin qubits \cite{burkardSemiconductorSpinQubits2023}. Such systems could incorporate concepts from quantum CCD architectures \cite{pinoDemonstrationTrappedionQuantum2021}, trapped-ion processors, and superconducting quantum technologies \cite{osada2022feasibility}.

An alternative route to electron confinement is provided by Penning traps\cite{brown1986geonium}. However, the strong magnetic fields\cite{jain_penning_2024} typically required for stable operation are incompatible with many of the optical transitions and Rydberg-state manipulations envisioned for hybrid electron-ion systems. For this reason, the present work focuses on an RF-based trapping architecture.

Despite these prospects, the experimental realization of dual-frequency trapping remains challenging. While theoretical studies predict broad stability regions for coupled electron-ion traps \cite{foot2018two,leefer2017investigation}, experimental implementations have demonstrated that stable operation is highly sensitive to electric-field imperfections. Electron confinement at GHz frequencies requires radio-frequency fields with high spatial homogeneity and stationary phase. Variations of phase distribution and anharmonicities lead to excess micromotion, enhanced heating, and particle loss \cite{berkeland2002minimization}. In practical devices, especially those based on printed-circuit-board electrodes, such imperfections arise from finite electrode dimensions, fabrication tolerances, surface roughness, and assembly inaccuracies.

To maintain the phase stability of the oscillating electric field over time and across space, all electron Paul traps integrate the trap into microwave resonators. Such a solution also allows achieving the high field amplitudes required for electron trapping while reducing dissipative heating. Superconducting resonators can provide exceptional quality factors and low-noise operation, enabling strong coupling to microwave circuitry and sensitive particle detection \cite{osada2022feasibility,taniguchi2025image}. However, these approaches generally rely on cavity-like geometries that restrict optical access to the trapping region. On the other hand, open geometries suitable for laser cooling, imaging, and manipulation of trapped ions typically exhibit lower quality factors and reduced field uniformity, limiting their applicability to electron confinement and dual-frequency operation.

Table~\ref{tab:comparison} summarizes the limitations of existing approaches and highlights the longstanding trade-off between field quality and experimental accessibility. To date, no platform has simultaneously provided the GHz trapping fields required for electrons, sufficiently homogeneous electric-field distributions, and the optical access needed for the preparation, manipulation, and imaging of trapped ions. Achieving this combination of properties requires a resonator geometry that combines efficient microwave confinement with large apertures for optical and particle-beam access, placing stringent demands on the design and fabrication process.

The realization of such resonator structures presents significant manufacturing challenges because field quality is highly sensitive to geometric deviations and surface properties. Additive manufacturing offers a route to producing such structures with reduced fabrication complexity and design constraints. However, its suitability for GHz resonators remains uncertain because surface roughness, dimensional tolerances, and material properties may significantly degrade resonator performance.

In this work, we present a resonator-based Paul trap architecture designed to address these competing requirements. The proposed device operates in the GHz regime necessary for electron confinement while retaining an open geometry compatible with laser access and imaging of ion Coulomb crystals. By combining moderate quality factors with a well-defined resonant field structure, it enables high electric-field amplitudes at reduced input power and mitigates microwave-induced heating while preserving field stability throughout the trapping region. We characterize the electromagnetic properties of the trap, evaluate its suitability for dual-frequency operation, and assess its compatibility with the confinement and long-term storage of electron spin qubits.

\begin{table*}[t]
\caption{Comparison of representative electron and ion trapping platforms. The table highlights the trade-off between electromagnetic performance and experimental accessibility addressed in this work. $^*$Orders of magnitude of the driving frequencies are given.}
\label{tab:comparison}
\begin{ruledtabular}
\begin{tabular}{lccccc}
System & Ref. & Frequency$^*$ & Quality factor & Optical access & Coupling with ions \\
\hline
Superconducting resonator trap & \cite{osada2022feasibility}& 1 GHz & very high ($>10^4$) & limited & restricted by design \\
PCB / planar GHz traps & \cite{matthiesen2021trapping,mikhailovskiiTrappingElectrons40Ca2026} & 1 GHz & low ($\sim 10$--$50$) & good & limited by instability \\
Coaxial resonator ion traps & \cite{jefferts1995coaxial} & 100 MHz & moderate ($\sim 10^3$) & moderate & not applicable \\
\textbf{This work} & -- & 1 GHz & moderate ($\sim 10^3$) & good & designed for cotrapping \\
\end{tabular}
\end{ruledtabular}
\end{table*}

\section{Implementation of Microwave Resonator Paul Trap}
\label{Proposed Design of Resonator}

The microwave resonator Paul trap was briefly introduced by \citet{kumar3DprintedComponentsElectronion2025a}, where the compatibility of the manufacturing method (laser powder bed fusion, L-PBF) with ultra-high vacuum operation and the realization of the microwave resonance were demonstrated. Here, we provide a detailed description of the device architecture, the properties arising from the manufacturing process and assess the reproducibility of the device in computer simulations.

\subsection{Design Requirements}

The requirements imposed by electron confinement differ substantially from those encountered in conventional ion traps. Owing to the small mass ratio \(m_{\mathrm e}/m_{\mathrm{Ca}}\sim10^{-5}\), stable electron confinement requires trapping frequencies in the GHz range if the characteristic dimensions of the trap are on the order of \qty{1}{mm} or less -- the range common to trapped-ion quantum computers\cite{hongNumericalInvestigationSegmentedblade2022,xu3DprintedMicroIon2025a,menonMonolithicSegmented3D2026}. At such frequencies, maintaining a spatially uniform RF phase throughout the trapping volume becomes a critical design requirement, since uneven phase distributions generate distortions of the pseudopotential that lead to excess micromotion and particle heating \cite{berkeland2002minimization}.

For an electrode-driven quadrupole trap with electrode length \(L_{\mathrm Q}\), the phase shift of the driving field is

\begin{equation}
  \phi = \frac{2\pi L_\text{Q}}{\lambda_{\mathrm{eff}}}
       = \frac{\Omega L_\text{Q}}{v_{\mathrm p}},
\end{equation}
where \(v_{\mathrm p}\) is the phase velocity of the RF signal and \(\lambda_{\mathrm{eff}} = 2\pi v_{\mathrm p}/\Omega\) is the effective wavelength along the conductor. At \qty{2.2}{GHz}, assuming \(v_{\mathrm p}\approx c\), phase shifts of approximately \(79^\circ\) and \(132^\circ\) arise over electrode lengths of \qty{30}{mm} and \qty{50}{mm}, respectively. Such variations are sufficiently large to compromise stable electron confinement.

In addition to phase stability, the trapping field must exhibit good spectral purity. Noise or parasitic signals near the electron secular frequencies \(\omega_i\) may couple energy into the motion and reduce confinement times. For this reason, the resonator bandwidth must remain sufficiently narrow that the trapping resonance is well separated from the secular-frequency spectrum.

These requirements motivate the incorporation of trap electrodes into resonant structures as demonstrated by \citet{osada2022feasibility,matthiesen2021trapping,mikhailovskiiTrappingElectrons40Ca2026}. Because the trapping field is generated as a standing-wave resonance, the phase remains nearly constant throughout the trapping region despite GHz operation. At the same time, resonant enhancement enables large electric-field amplitudes to be achieved with reduced input power, thereby minimizing dissipative heating. The resonator also acts as a frequency-selective filter, suppressing unwanted frequency components. 

\subsection{Design and Its Realization}
\label{sec:relization}

\begin{figure*}
    \centering
    \includegraphics[width=0.9\textwidth]{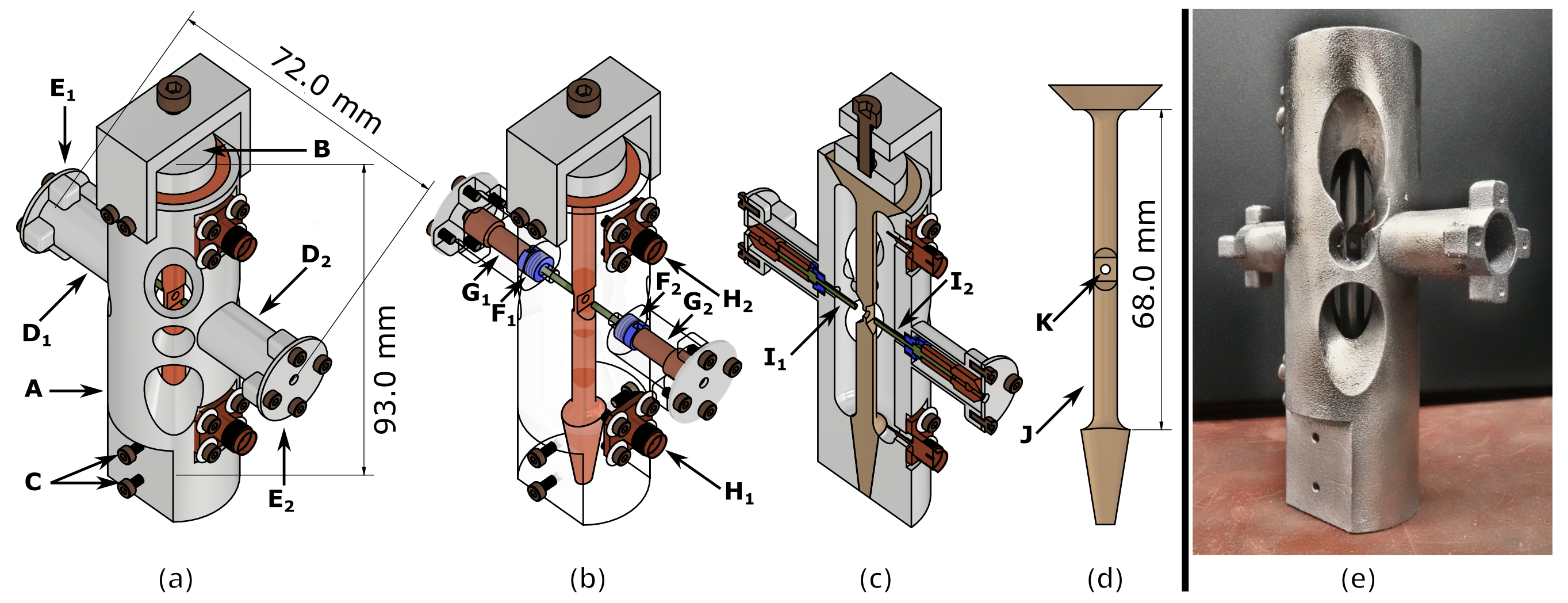}
\caption{Illustration of the EiTEx 3D-printed resonator. \textbf{a)} The complete assembled design shows the cavity (\textbf{A}) housing the resonator rod, which is press-fitted using (\textbf{B}) and can be mounted to the chamber blank flange with screws (\textbf{C}). Perpendicular extensions (\textbf{D}$_{1,2}$) hold the endcaps, with cover plates (\textbf{E}$_{1,2}$). \textbf{b)} Internal view showing the resonator rod and opposing endcaps aligned with the rod hole. Ion trapping RF signal is supplied via inline connectors (\textbf{G}$_{1,2}$), while electron-trapping signals are delivered through connectors (\textbf{H}$_{1,2}$). \textbf{c)} Ion trapping signal is delivered to endcaps (\textbf{I}$_{1,2}$). \textbf{d)} Electron trapping signal is coupled to a resonating rod (\textbf{J}), which includes a hole (\textbf{K}) for trapping. \textbf{e)} Photograph of a manufactured trap.}
\label{fig:1}
\end{figure*}

Our design of the trap is illustrated in Figure~\ref{fig:1}. The assembly consists of an outer cavity body (\textbf{A}, Figure~\ref{fig:1}a containing a central resonator rod (\textbf{J}), which is held in position by a press-fit mount (\textbf{B}) and attached to a vacuum flange using screws (\textbf{C}). Reduction of the main constituting parts to two, \textbf{A} and \textbf{J}, allows us to avoid reductions in the resonance quality factor arising from imperfect interfaces between assembled parts and from disruptions in the continuity of the cavity’s concave surfaces.

Two perpendicular extensions (\textbf{D}$_{1,2}$) protrude from the cavity and support the ion-trap endcaps (\textbf{I}$_{1,2}$), which are enclosed by cover plates (\textbf{E}$_{1,2}$). An internal view of the assembled structure is shown in Figure~\ref{fig:1}b, where the endcaps are aligned with an aperture in the resonator rod to define a common circular trapping region (\textbf{K}) with a diameter of \qty{2}{mm} at the center of the device.

When driven at a microwave frequency \(\Omega/2\pi\), the resonator supports standing radio-frequency modes along its longitudinal (\(x\)) axis. For the present geometry, the electromagnetic field is dominated by the fundamental half-wavelength mode, producing the electric-field distribution required for electron confinement in the central trapping region. The microwave signals for electron trapping are supplied through the connector (\textbf{H}$_{1}$), while the radio-frequency signals for the confinement of ions are delivered independently to the endcaps through inline connectors (\textbf{G}$_{1,2}$). Driving the endcaps at frequency \(\Omega_{\mathrm{I}}/2\pi\) generates the ion-trapping pseudopotential, whereas the resonant microwave field provides confinement of electrons. The device therefore combines both trapping mechanisms within a single structure and operates as a dual-frequency Paul trap\cite{foot2018two}.

The final trap shown in Figure \ref{fig:1}b was manufactured from the AlSi10Mg alloy using the L-PBF additive process and subsequently polished. The uncertainty of the dimensions of the components accumulated during the manufacturing process is estimated to be around \qty{100}{\micro\metre} (Appendix \ref{sec:dimensionerror}), which worsens the contact between parts \textbf{A} and \textbf{J} crucial to maintaining the quality of the resonator. The polishing reduces the root-mean-square roughness of the printout from $R_\mathrm{q} = \qty{14\pm10}{\micro\metre}$, to \qty{2.5\pm0.7}{\micro\metre} which is comparable to the skin depth of approximately \qty{2}{\micro\metre} for \qty{2}{GHz} signals assuming the material composition has not changed in the printing process. The measurement of the surface roughness and the analysis of material composition that allow us to state these figures authoritatively are described in Appendix \ref{sec:surfaceroughness} and \ref{sec:composition}. 

\begin{figure*}
    \centering
    \includegraphics[width=15cm]{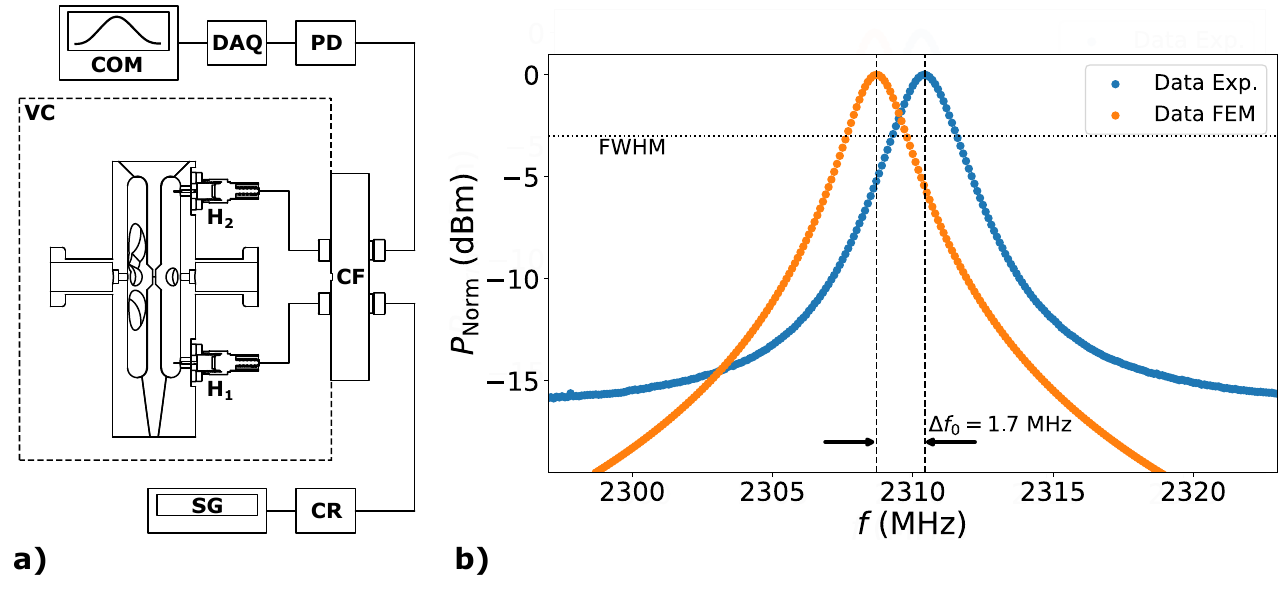}

    \caption{\textbf{a)} Experimental scheme for measuring resonance behaviour. The signal generated by programmable signal generator (\textbf{SG}) is fed through a circulator-based isolator (\textbf{CR}) and vacuum feedhthrough on a conflat flange (\textbf{CF}) to a pin of connector \textbf{H}$_1$ situated inside the vacuum chamber (\textbf{VC}) near the node of the oscillating field. The signal in the cavity is measured through the mirrored connector \textbf{H}$_2$ by a power detector (\textbf{PD}) and digitized by a data acquisition board (\textbf{DAQ}). The data are then processed by a computer (\textbf{COM}) running a signal analysis program, which controls the signal generator SG. The models of devices are described in the work by \citet{kumar3DprintedComponentsElectronion2025a}. \textbf{b)} Comparison of FEM simulated and experimentally measured resonance frequencies. A line is plotted to mark the location of the Full Width at Half Maximum (FWHM). $\Delta f_0$ denotes the difference between the resonance frequencies obtained from simulation and one measured experimentally. The powers have been normalized to result in the peaks to be at 0 dBm.}
    \label{fig:resonances}
\end{figure*}

The roughness, together with aforementioned mismatching between parts, was threatening to reduce the resonance quality factor of the resonator; therefore, the resonance curve of the resonator (without endcaps inserted) was measured (as depicted in Figure \ref{fig:resonances}a) and compared with the simulations\cite{artiq}  carried out in COMSOL Multiphysics\textsuperscript{\textregistered} with the temperature variable set to \qty{20}{\celsius} (laboratory temperature) and the pressure set to \qty{1e-9}{Pa} (equals approximately \qty{1e-11}{mbar}), even though two orders higher pressure was also found to have negligible effect on results. The simulations were carried out by using COMSOL interface "Electromagnetic Fields, Frequency Domain" under RF Module. The used mesh was pre-defined coarse mesh outside the design and pre-defined fine mesh in most parts of the design. The trapping region was modelled using an even finer, custom mesh, which allowed elements down to around 20 \textmu m in size to be formed.

The comparison of the model with the measurement (further described in Appendix \ref{sec:verification}), shown in Figure \ref{fig:resonances}b, indicates that the manufacturing imperfections did not have an overly detrimental effect. The resonance frequency is at \qty{2.3}{GHz} and the mismatch between the simulated and measured values is around \qty{2}{MHz}. From the simulated and measured spectra, we obtain quality factors \num{1076\pm9} and \num{990\pm10}, respectively. The signals in the \qtyrange{1}{100}{MHz} range, the expected range of electrons' secular motion frequency, are  attenuated by roughly four orders of magnitude.

\section{Electromagnetic Features}
\label{sec:elmagfeatures}

The comparison of COMSOL simulation results with the physical measurement of the resonance indicates that the simulations reasonably predict reality. Therefore, we can now use the same ``virtual device'' to study what the fields inside the trapping volume look like and predict the electron trapping capabilities.

 Maxwell’s equations were solved for a \(10\,\mathrm{W}\) microwave signal injected through \(\mathbf{H}_1\), the resonator treated as electrically floating so that the field distribution is determined by its natural cavity modes. 

 \begin{figure*}
    \centering
    \includegraphics[width=1.0\textwidth]{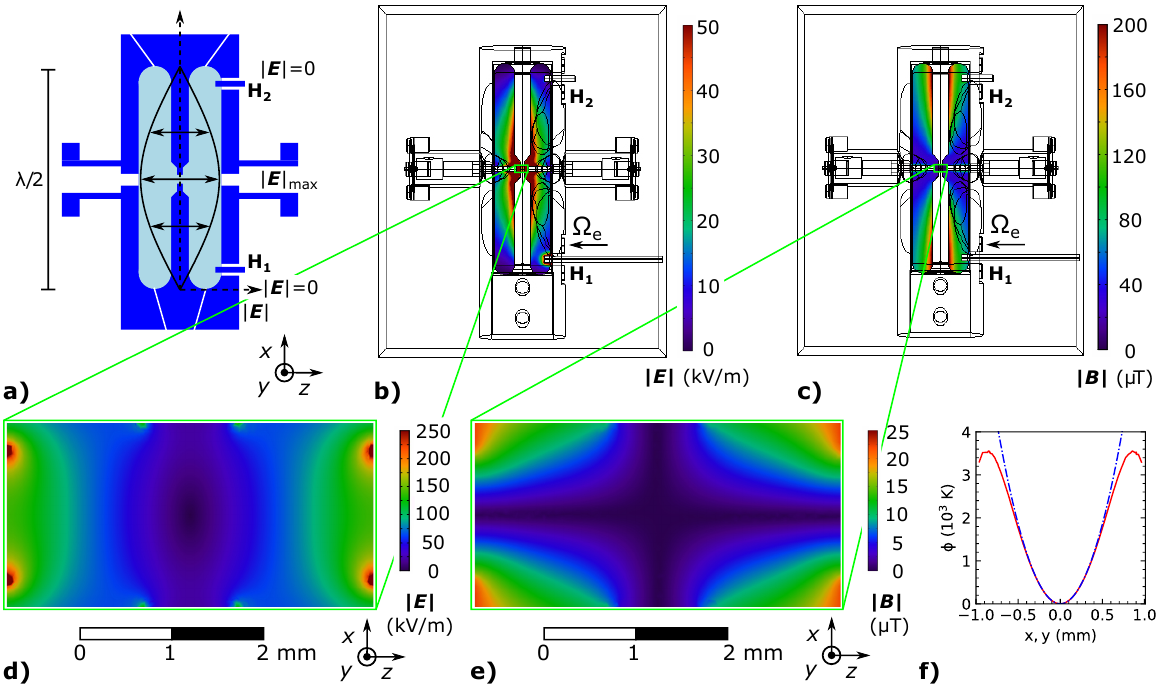}
\vspace{-2.5em} % Adjust the negative value as needed to reduce the gap
\caption{Simulated field distribution of the resonator. \textbf{a)} Illustration of resonance mechanism. The endcap electrodes are not depicted for clarity. The fundamental TEM mode (at \qty{2.21}{GHz} in the case of final assembly) forms a strong, well-confined field along the central axis, excited from the bottom SMA jack. \textbf{b)} The COMSOL simulation result for electric field amplitude confirms the field shape shown in the schematic. \textbf{c)} Simulated distribution for the magnetic field amplitude. Unlike the electric field, its amplitude along the central axis has a minimum in the middle of the resonator. \textbf{d)} and \textbf{e)} are showing details of the distributions in \textbf{b)} and \textbf{c)}, respectively. The field ranges are different for full images and insets, in order to improve the visibility of the dominant field features at each volume. \textbf{f)} Pseudopotential for electron trapping along the $x$- or $y$-axis (solid red), compared with an ideal harmonic potential (blue dash-dotted). The lower black dotted curve shows the deviation between them, highlighting non-idealities in the trapping system.}
\label{fig:2}
\end{figure*}

 Simulations indicate that the resonant mode of the trap with endcap electrodes installed occurs at approximately \(\Omega\sim 2.21\,\mathrm{GHz}\) -- the electromagnetic field distribution is depicted in Figures~\ref{fig:2}b,d. The fundamental half-wavelength mode (\(\lambda/2\), n=1) dominates the distribution of electric field $\mathbf{E}$, while higher-order modes \((n>1)\) are strongly suppressed \((E_\text{n}/E_1 \approx 10^{-3})\). As a result, a well-defined trapping field is formed near the center of the resonator (\(\textbf{K}\)). The calculated quality factor for this mode is \(\sim 1050\).

Figure~\ref{fig:2}f shows the trap pseudopotential derived from the simulated electric-field distribution \(\mathbf{E}\), \[ \Psi(x,y,z)=\frac{q^{2}|\mathbf{E}|^{2}}{4\me\Omega^{2}}, \] with a maximum depth of \(\phi=\Psi_{\max}/k_B\approx \qty{3500}{K}\). A least-squares fit of the potential near the trap center \(\mathbf{r}_0\) along the principal axes, \(\delta r_i=(r_i-r_{0,i})\) (where $\delta r_x \equiv x$, $\delta r_y \equiv y$, $\delta r_z \equiv z$), to 
\begin{equation}
\label{eq:elpot}
\Psi_i = \frac{\me\omega_i^2}{2}\delta r_i^2 + \beta_i \delta r_i^4 + \mathcal{O}(\delta r_i^6), 
\end{equation}
gives the secular frequencies \(\omega_x=\omega_y=\qty{77(2)}{MHz}\) and \(\omega_z=\qty{154(3)}{MHz}\). Within the stated error bounds, these frequencies are consistent with those derived from the Fourier analysis of electron motion, as computed by the finite-difference time-domain method using the same simulation software. For the anharmonicity parameters, we obtain $\beta_x = \qty{11.8(0.6)}{\joule\per\milli\metre\tothe{4}}$, $\beta_y = \qty{7.6(0.7)}{\joule\per\milli\metre\tothe{4}}$ and $\beta_z=\qty{79(35)}{\joule\per\milli\metre\tothe{4}}$. The large errors for these values are given by the fact that the electric field is not perfectly symmetric. From the value of $\beta_z$, we obtain $K_z \approx \hbar \times \qty{0.9(5)}{\per\second}$ as the Kerr parameter appearing in the motional Hamiltonian of the trapped-electron system,
\begin{equation}
\label{eq:hm}
H_\mathrm{m} = \sum_{i=x,y,z}\big[\hbar \omega_i + K_i \hat{n}_i\big]\hat{n}_i + \text{const.},
\end{equation}
where $\hat{n}_i $ denotes the number operator $(a_i^\dagger a_i)$. Based on a sixth-order fit for the pseudopotential of the trap, its second derivative remains positive up to a 0.56 mm distance from the centre of the trap, In this region, the trap can accommodate $6\cdot 10^5$ motional quanta with an energy difference $\hbar\omega_z$. Even at such a high excitation level, the anharmonic term contributes to the total energy by around 8\%.

Figures \ref{fig:2}c,e illustrate the distribution of the magnetic field amplitude $B$. We find that the magnetic field induced by the oscillating electric field can be approximated as
\begin{equation}
\label{eq:magfield}
\mathbf{B}(\mathbf{r},t) =  (\mathbf{B}_{\mathrm{0,RF}}+\mathsf{G}\cdot\delta\mathbf{r})\cos{(\Omega t)} + \mathcal{O}(\delta\mathbf{r}^2), 
\end{equation}
where $\mathsf{G}$ is the magnetic-field gradient tensor with components 
\begin{equation} 
G_{ij} = \frac{\partial B_i}{\partial r_j} \,,
\end{equation}
and $\mathbf{B}_{0,\mathrm{RF}}$ is the constant oscillating field in the trap centre.

For these quantities -- that are determined by the trap geometry -- we obtain values
\begin{align*}
    \mathbf{B}_{\mathrm{0,RF}}&=\left(\num{151(3)},\num{-87(3)},\num{1(1)}\right)\unit{\nano\tesla}\\
    \mathsf{G}&=\begin{pmatrix}
        \num{12(2)}&\num{7(2)}&\num{-47(2)}\\
        \num{12(2)}&\num{-21(2)}&\num{-53(3)}\\
        \num{-50(2)}&\num{-55(2)}&\num{5(2)}
    \end{pmatrix}\unit{\micro\tesla\per\metre}.
\end{align*}

We see that the gradient matrix has the symmetricity perturbed in the $xy$ plane while it is preserved within error margins in the $xz$ and $yz$ planes. This could be a result of asymmetric placement of openings for optical and particle beams.

\section{Investigation of compatibility with electron-spin qubit storage}
\label{sec:spinok} 

With the spatial profile of the oscillating magnetic field established, a natural question follows: Can the trap enable the storage of quantum information in the electron spin? Since the motional degree of freedom is coupled to the spin through a magnetic-field gradient, motional heating directly translates into qubit dephasing.

The electron spin Hamiltonian is 
\begin{equation*} 
H_\mathrm{s} = \frac{\gamma_{\mathrm{e}}}{2} \boldsymbol{\sigma}\cdot\mathbf{B}(\mathbf{r}), 
\end{equation*} 
where $\gamma_\text{e}$ is the electron's gyromagnetic ratio. To the field \eqref{eq:magfield}, we superimpose a static quantization field $\mathbf{B}_0=(0,0,B_0)$ and obtain the Hamiltonian
\begin{equation*}
H_\mathrm{s} = \hbar \frac{\gamma_{\mathrm{e}}}{2}\boldsymbol{\sigma}\cdot \mathbf{B}_0 + \hbar\frac{\gamma_{\mathrm{e}}}{4}\boldsymbol{\sigma}\cdot (\mathbf{B}_{0,\mathrm{RF}}+\mathsf{G}\cdot\delta\mathbf{r})(e^{i\Omega t}+e^{-i\Omega t}),    
\end{equation*}
 where  only the lowest-order position dependence is retained. The first term simplifies to $\hbar (\gamma_\mathrm{e}/2) \sigma_z B_0$. The second term describes the spin-motion coupling. 

 The high-frequency approximation in terms of $\Omega$ ($\gg \omega_{\mathrm{s0}},\, \omega_z$)\cite{goldmanPeriodicallyDrivenQuantum2014} yields an effective spin Hamiltonian
 \begin{equation*}
 \begin{split}
     H_{\mathrm{s,eff}} = &\frac{\hbar \omega_{\mathrm{s}0}}{2}\sigma_z \\&-\frac{\hbar \gamma_{\mathrm{e}}^2 \omega_{\mathrm{s}0}}{8 \Omega^2} [(g_x^2 + g_y^2)\sigma_z-g_z(g_x \sigma_x + g_y \sigma_y)] \\&+ \mathcal{O}(\Omega^{-3}) \,,
     \end{split}
 \end{equation*}
 where $\omega_{\mathrm{s}0} = \gamma_\mathrm{e}B_0$ and 
\begin{equation*}
    g_i \equiv B_{0,\mathrm{RF},i}+\sum_j G_{ij}\delta r_{0,j}(a_j + a^\dagger_j).
\end{equation*}

Performing rotating wave approximation in $\omega_{\mathrm{s}0}$ rotating reference, we obtain the spin Hamiltonian projected to spin quantization axis
\begin{equation*}
    H_{\mathrm{s},\parallel} = \lambda(g_x^2+g_y^2)\sigma_z,\qquad\lambda=-\frac{\hbar \gamma_{\text{e}}^2 \omega_{\mathrm{s}0}}{8 \Omega^2}\,.
\end{equation*}

Neglecting rapidly oscillating terms (including the cyclotron motion of the electron), the total effective Hamiltonian for a single axial mode becomes
\begin{equation*}
    H_{\mathrm{eff}} = H_0 + \lambda F\sigma_z,
\end{equation*}
where $H_0$ and $F$ are 
 \begin{eqnarray*}
H_0 &=& \hbar \omega_z \hat{n}_z + K_z \hat{n}_z^2,\\
    F &=& F_0 + F_1(a_z+a_z^\dagger) + F_2 \hat{n}_z 
\end{eqnarray*}
with 
\begin{eqnarray*}
F_0 &=& |\mathbf{B}_\perp|^2+|\mathbf{G}_\perp|^2 \delta z_0^2 \\
        F_1 &=& 2(\mathbf{B}_\perp \cdot \mathbf{G}_\perp)\delta z_0\\
        F_2 &=& 2|\mathbf{G}_\perp|^2 \delta z_0^2\\
    \mathbf{B}_\perp &=& (B_{0,\mathrm{RF},x},B_{0,\mathrm{RF},y})\\
    \mathbf{G}_\perp &=& (G_{xz},G_{yz})\,.
\end{eqnarray*}
In this way, we construct a pure-dephasing spin–boson model\cite{schlosshauerQuantumDecoherence2019} with a Kerr-type nonlinearity in the mode\cite{lorchGenuineQuantumSignatures2016}.

We assume that the electron is initially prepared in a low-occupation Fock state $\ket{n_0}$ and subsequently undergoes heating due to electric-field noise with the rate  
\begin{equation}
\label{eq:hrate}
\begin{split}
    \Gamma(n_z) \approx &\frac{\qe^{2}}{4\me\hbar\omega_{i,n_i}}S_{E}(\omega_{z,n_z})\\
    \approx &\frac{\qe^2}{4\me\hbar\omega_z} S_E + \mathcal{O}(K_z n_z),%\frac{\qe^2 S_E}{4\me}\left[\frac{1}{\hbar \omega_z}-\frac{(2 n_z+ 1) K_z }{(\hbar \omega_z)^2}\right.\\
    %& \left.+ \frac{(2n+1)^2 K_z^2}{(\hbar \omega_z)^3}\right]+\mathcal{O}(K_z^3 n_z^3)
\end{split}
\end{equation}
where $S_E$ is a spectral density of noise\cite{brownnuttIontrapMeasurementsElectricfield2015}, which we approximate as a constant in the relevant range of frequencies $\omega_{z,n_z}$. As $\omega_z \simeq \omega_{z,n_z}$, we approximate $\Gamma$ by the constant term in the description from here on.

At room temperature, the Fock states get populated at the level of
\begin{equation*} 
n_\mathrm{th} = \frac{\kB T}{\hbar\omega_z} \gtrsim 10^4 
\end{equation*} for $\omega_z = 2\pi\times\qtyrange{10}{100}{MHz}$. Since $n_0\ll n_{\mathrm{th}}$, the thermal occupation remains much larger than the motional occupation throughout the relevant dynamics. The evolution can therefore be approximated by the infinite-temperature master equation
\begin{equation*}
    \frac{d\rho}{dt}=-\frac{i}{\hbar}[H_{\mathrm{eff}},\rho]+\Gamma(\mathcal{D}[a^\dagger]+\mathcal{D}[a])\rho\,,
\end{equation*}
where $\mathcal{D}[L]\rho=L\rho L^{\dagger}-(1/2)\{L^{\dagger}L,\rho\}$.

The coherence dynamics are described by a Lindbladian for the off-diagonal spin density matrix element\cite{trushechkinQuantumMasterEquations2022a} in the form
\begin{equation}
\label{eq:mastereq}
\begin{split}
\frac{d\rho_{\uparrow\downarrow}}{dt} =&  -\frac{i}{\hbar} [H_0,\rho_{\uparrow\downarrow}] - \frac{i\lambda}{\hbar}\{F,\rho_{\uparrow\downarrow}\} \\ 
&+\Gamma\left[\mathcal{D}\!(a_z)+\mathcal{D}\!(a_z^\dagger)\right]\rho_{\uparrow\downarrow}.\\
\end{split}
\end{equation}

Moving into the $\hbar \omega_z n$ interaction picture,
\begin{equation*}
    \frac{d\tilde{\rho}_{\uparrow\downarrow}}{dt} = (\mathcal{L}_0 + \mathcal{L}_{-1}e^{-i\omega_z t} + \mathcal{L}_{+1}e^{+i\omega_z t})\tilde{\rho}_{\uparrow\downarrow} 
\end{equation*}
where 
\begin{align*}
\mathcal{L}_0 &\begin{multlined}[t]
    = -\frac{i K_z}{\hbar}[ \hat{n}_z^2,\cdot] -\frac{i\omega_z}{2}\{\eta_0 + \eta_2 \hat{n}_z, \cdot \} \\+ \Gamma\left[\mathcal{D}\!(a_z)+\mathcal{D}(a_z^\dagger)\right]
\end{multlined}\\
    \mathcal{L}_{-1} &= -\frac{i\omega_z}{2}\eta_1\{a_z, \cdot \}\\
    \mathcal{L}_{+1} &= -\frac{i\omega_z}{2}\eta_1\{a_z^\dagger, \cdot \}
\end{align*}
with dimension-less factors
\begin{equation*}
    \eta_{0,1,2} = \frac{2\lambda F_{0,1,2}}{\hbar \omega_z}.
\end{equation*}

The van Vleck high-frequency expansion\cite{ikedaNonequilibriumSteadyStates2021} of this -- carried out assuming $\hbar \omega_z \gg K_z n_z$  and $\omega_z \gg \lambda$ -- leads to an effective Liouvillian
\begin{equation}
\label{eq:effl}
    \tilde{\mathcal{L}} = \mathcal{L}_0-i\frac{K_z}{\hbar }\eta_1^2\left[\hat{n}_z,\cdot\right]  +\eta_1^2\left(\Gamma-i\frac{\omega_z}{4} \eta_2\right)\,\mathds{1}+ \mathcal{O}(\omega_z^{-3})\,
\end{equation}
that is applied to a transformed density operator $\tilde{\rho}_{\uparrow\downarrow}$. 

This Liouvillian preserves the difference $m-n$ between Fock-state indices. Hence the diagonal sector is closed under the dynamics. Starting from a state localized in a single Fock state,
\[
\tilde{\rho}_{\uparrow\downarrow}(0)=|n_0\rangle\langle n_0|,
\]
the transformed density operator remains diagonal for all times,
\begin{equation}
\label{eq:tilderho}
\tilde{\rho}_{\uparrow\downarrow}(t)=\sum_{n=0}^{\infty} c_n(t)\,|n\rangle\langle n|.
\end{equation}

The spin coherence is given by the trace of the off-diagonal spin block,
\begin{equation*}
C(t)=\operatorname{Tr}\!\left[\tilde{\rho}_{\uparrow\downarrow}(t)\right]
=\sum_{n=0}^{\infty} c_n(t).
\end{equation*}

Substituting Eq.~\eqref{eq:tilderho} into the master equation generated by the effective Liouvillian \eqref{eq:effl}, and using that the diagonal sector is invariant under the dynamics, we obtain
\begin{equation*}
\begin{split}
\frac{d c_n}{dt}
={}&\Gamma(n+1)c_{n+1}
+\Gamma n\,c_{n-1}\\
&+\Bigl[
-\Gamma(2n+1)
-i\omega_z(\eta_0+\eta_2 n)\\
&+\eta_1^2\Bigl(\Gamma-i\frac{\omega_z}{4}\eta_2\Bigr)
\Bigr]c_n ,
\end{split}
\end{equation*}
with $c_{-1}(t)\equiv 0$.
Equivalently,
\begin{equation*}
\begin{split}
\frac{d c_n}{dt}
={}&\Gamma(n+1)c_{n+1}
+\Gamma n\,c_{n-1}\\
&-\bigl(2\Gamma+i\omega_z\eta_2\bigr)n\,c_n\\
&-\left[
\Gamma(1-\eta_1^2)+i\omega_z\left(\eta_0
+\frac{\eta_1^2 \eta_2}{4}\right)
\right]c_n .
\end{split}
\end{equation*}

Solution of this equation by the method of characteristics is described in Appendix \ref{sec:coherencederiv}. For the coherence, we obtain 
\begin{equation}
\label{eq:coherence}
\begin{split}
\ln C(t)=
\left[
\Gamma\eta_1^2
+i\omega_z\left(\frac{\eta_2}{2}-\eta_0-\frac{\eta_1^2\eta_2}{4}\right)
+\Gamma s
\right]t\\
+n_0\ln\!\left(\frac{r+e^{2\Gamma s t}}{r+1}\right)
-(n_0+1)\ln\!\left(\frac{1+r\,e^{2\Gamma s t}}{r+1}\right)
\end{split}
\end{equation}
with
\begin{equation*}
r=1+\frac{i\omega_z\eta_2}{2\Gamma}+s,
\qquad
s=\sqrt{\left(1+\frac{i\omega_z\eta_2}{2\Gamma}\right)^2-1}.
\end{equation*}

The square root is understood on the branch satisfying $\mathrm{Re}(s)\le0$. This choice follows from continuity with the uncoupled limit $\eta_{0,1,2}\rightarrow0$ and ensures that the coherence $C(t)$ remains bounded for $t\rightarrow\infty$. The opposite branch corresponds to exponentially growing characteristic solutions and yields an unphysical long-time behaviour of the coherence.

We define coherence time as $\tau$, such that satisfies the relation \[ \mathrm{Re}[\ln C(\tau)] = -1\,.\]

The largest values appearing in the \eqref{eq:coherence} are the initial occupation number $n_0$ and the heating rate $\Gamma$. We will discuss their values in the next subsections.

\subsection{Determination of Heating Rate}
\label{sec:hrate}

The heating rate $\Gamma$ depends on the spectral noise density of electric field interacting with the trapped-electron quantum harmonic oscillator. Johnson–Nyquist noise constitutes the fundamental thermal electric-field-noise floor of metallic trap electrodes. While additional surface-induced mechanisms often dominate motional heating in miniaturized ion traps, Johnson noise provides the baseline contribution against which other sources are commonly compared\cite{brownnuttIontrapMeasurementsElectricfield2015}.

For trap with electrodes situated at the distance of $d_\mathrm{e}$ at temperature $T_\mathrm{r}$, we get the noise spectral density\cite{brownnuttIontrapMeasurementsElectricfield2015}
\begin{equation}
\label{eq:se}
    S_E = \frac{4 k_\mathrm{B}T_\mathrm{r} R(\omega_z,T_\mathrm{r})}{d_\mathrm{e}^2}\,,
\end{equation}
where $k_\mathrm{B}$ is the Boltzmann constant and $R$ is the electrode material resistance as a function of frequency and temperature.

The resistance $R$ of a material with root-mean-square roughness $R_\mathrm{q}$ is derived from the resistance of an ideal smooth surface $R_
\mathrm{id}$ according to the Hammerstad model \cite{abdelatty2023role}:
\begin{equation}
\label{eq:resistance}
R = R_\mathrm{id} K_\text{sr} = R_\mathrm{id}\left(1 + \frac{2}{\pi} \arctan\left[1.4 \left(\frac{R_\text{q}}{\delta}\right)^2\right]\right),
\end{equation}
where \(\delta=\sqrt{\frac{2\rho}{\omega\mu}}\) is the skin depth calculated from the resistivity $\rho$ at frequency $\omega_z$ with $\mu$ being the permeability of aluminum, which is approximately equal to the vacuum permeability.

That clearly indicates that the heating rate $\Gamma$ depends not only on the design but also on the material parameters such as root-mean-square surface roughness and the resistivity, the values of which were given in Section \ref{sec:relization}. Assembling these variables in Equations \eqref{eq:hrate} and \eqref{eq:se}, and taking the uncertainties of the variables into account, we get $\Gamma = \qty{101\pm24}{\per\second}$  for the secular frequency $\omega_z = 2\pi\times\qty{154\pm3}{MHz}$. 

\subsection{Choice of Initial Motional State}
\label{sec:initialstate}

The methods for preparing initial motional state $\ket{n_0}$ are multiple. One of them would be a cryogenic cooling as suggested by \citet{osada2022feasibility}. Such a method will extend coherence times to significantly more than hours, allowing convenient manipulation of the qubits. However, that comes with the expense of incorporating a complex and power-hungry cryogenic system, making the scalability of such an architecture questionable.

Another approach of yielding a pure state $\ket{n_0}$ would be to detach the electron coherently from the laser-cooled ion stored in the centre of the trapping region. In such a case, the electron exists in a potential landscape carved by the Coulomb potential from the doubly charged ion and the Paul trap:
\begin{equation}
V(r) =  \frac{1}{2} m_\mathrm{e} \omega_\mathrm{e}^2 \delta r^2-\frac{2 e^2}{4\pi\epsilon_0 \delta r}.
\end{equation}

The exact method of reaching that condition can be based on the dual-frequency trapping mechanism\cite{foot2018two}, optionally complemented with optical trapping\cite{schaetzTrappingIonsAtoms2017}. The presented trap is equipped with large optical windows to allow testing such an implementation.

Naturally, the presence of the ion distorts the harmonicity of the field for the electron, driving the coherence time estimation from the previous section unusable. That is true for the low levels of excitation at which $|V(r)| < 0$. However, at sufficiently large excitation levels, the Coulomb field acts merely as a perturbation to the harmonic trapping field. That happens approximately above the excitation energies
\begin{equation*}
    E_c 
    = \frac{1}{2} m_e \omega_e^2 r_c^2\,,
\end{equation*}
where characteristic length scale
\begin{equation*}
    r_c = \left(\frac{Z e^2}{4\pi\epsilon_0 m_e \omega_e^2}\right)^{1/3}.
\end{equation*}

As far as $E_c/(\hbar\we)\equiv n_c \ll n_0$, the coherence time estimation is valid. For $\omega_z = 2\pi\times\qty{154}{MHz}$ and $Z=2$, $n_\mathrm{c} = 275$, therefore $n_0 = 10\times n_\mathrm{c}$ appears to be a safe choice.

\subsection{Sensitivity Analysis}
\label{sec:sensitivity}

With the expected values of parameters $n_0$ and $\Gamma$ on the table, we are ready to investigate the sensitivity of coherence time $\tau$ on their variation. 

Firstly, we evaluate the coherence time for our simulated system using the heating rate $\Gamma$ determined in Section \ref{sec:hrate}, and values of electric and magnetic field parameters derived from the model and given in Section \ref{sec:elmagfeatures}. During the derivation of the formula for the coherence time, several assumptions were made, and those are written in Table \ref{tab:validity}. To satisfy the condition for high frequency approximation, we take the quantization field $\mathbf{B}_0 = (0,0,10(1)\,)\unit{\milli\tesla}$, which results in $\omega_{s0}=\qty{1.7e9}{\per\second}$ -- larger than $\omega_z = 2\pi \times \qty{154}{MHz}$ and lower than the driving frequency $\Omega \approx 2\pi \times \qty{2}{GHz}$. To propagate all errors of fields described in Section \ref{sec:elmagfeatures}, we performed the Monte-Carlo calculation on 10000 samples and obtained $\tau= \qty{21\pm7}{\milli\second}$ for $n_0 = 10\times  n_c = 2750$, which is experimentally verifiable.

\begin{figure}
    \centering
    \includegraphics[width=\linewidth]{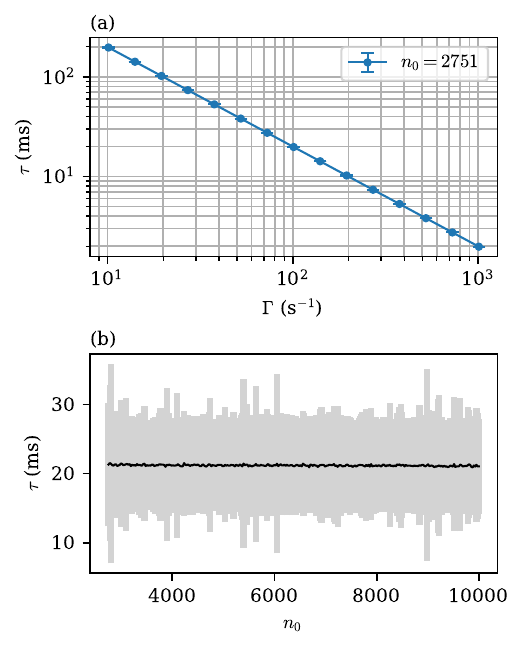}
    \caption{Dependence of coherence time $\tau$ on heating rate $\Gamma$ (a) and initial motional excitation $n_0$ (b). The dependence on $\Gamma$ was evaluated for $n_0$ being 10 times larger than the critical quantum number $n_c$. The size of error bars is smaller than the point size. The dependence on $n_0$ is burdened by large errors caused by the uncertainty of $\Gamma$.}
    \label{fig:tauscan}
\end{figure}

Keeping the aforementioned parameters as the baseline, we vary either the value of $\Gamma$ or $n_0$ and evaluate the coherence time using a Monte-Carlo method with the same number of samples. Figure \ref{fig:tauscan}a illustrates the dependence of the coherence time on the variation of heating rate $\Gamma$. As the value of  $\Gamma$ increases from \qtyrange{10}{1000}{\per\second}, $\tau$ follows a $1/\Gamma$ dependence and decreases from values over \qty{190}{ms} to \qty{2}{ms}. With the values of the heating rate fixed in each evaluation, the relative errors of the coherence times do not exceed the level of $10^{-3}$. This result illustrates that the span of $\Gamma$ values significantly exceeding the initial error estimates still results in practically verifiable coherence times. Changes in the coherence time with variation of $n_0$ (Figure \ref{fig:tauscan}b) are too small to outweigh the uncertainty introduced by the large error bars in the heating rate.

The comparison of the two dependencies indicates, that  -- in experimentally relevant scales -- the noise suppression is more important than the exact level the electron is excited to.

\begin{table}[]
    \centering
    \begin{tabular}{c|c}
    Description & Condition\\
    \hline
         Secular approximation& $K_i n_i \ll \hbar \omega_i$\\
         Harmonic field approximation & $n_c \ll n_0 $\\
         High frequency approximations & $\Omega\ \gg \omega_{\mathrm{s}0} \gg \omega_i$\\
         Weakness of magnetic fields & $|\eta_{0,1,2}| \ll 1$
    \end{tabular}
    \caption{Conditions on frequencies describing the system}
    \label{tab:validity}
\end{table}

\section{Discussion and Conclusion}
\label{Conclusions and Discussion}

While the sequential confinement of either electrons or ions has previously been demonstrated by \citet{}, the existing dual-frequency implementation, derived from an earlier single-frequency electron trap \cite{matthiesen2021trapping}, lacks the properties required for simultaneous confinement of both species. Furthermore, electron confinement times on the order of only \qty{1}{\milli\second} remain insufficient for many applications in quantum technologies. As suggested by \citet{matthiesen2021trapping}, this limited performance may originate from trap anharmonicities and from the instability of electron trajectories associated with electron creation away from the trap centre.

The trap architecture presented in this work was motivated by these limitations. As a countermeasure, we propose generating the electron through photoionization of a laser-cooled ion located at the trap centre. This approach eliminates trajectory instabilities associated with off-centre electron generation while simultaneously providing control over the initial motional state of the electron, as discussed in Sec.~\ref{sec:initialstate}. In the coherence analysis presented here, the influence of the parent ion is neglected once the electron occupies sufficiently high motional states, where the Coulomb interaction acts only as a weak perturbation of the trapping potential. While this approximation is supported by the energetic arguments of Sec.~\ref{sec:initialstate}, a rigorous quantum-mechanical treatment of the coupled electron-ion system remains an important subject for future work.

The requirement for laser cooling necessitates optical access to the trapping region and, therefore, the introduction of openings into the resonator body. Such modifications are expected to degrade the resonator performance and reduce the achievable trapping potential. To assess these effects, we measured the resonance spectrum of the fabricated resonator and found good agreement with numerical simulations. The validated model was subsequently used to determine the electric- and magnetic-field distributions within the trapping region. The resulting pseudopotential depth, exceeding \qty{3000}{K}, indicates that the device should support the proposed electron and hybrid electron-ion experiments.

The computer simulations helped us realize the presence of an inhomogeneous oscillating magnetic field generated intrinsically by the trapping fields. Although time-dependent magnetic-field gradients have previously been considered for electron-spin readout \cite{pengSpinReadoutTrapped2017}, to our knowledge, the influence of the magnetic field induced directly by the trapping field has not been investigated in detail. We therefore developed a model describing electron-spin decoherence in the present device and obtained a coherence time of approximately \qty{60}{\milli\second}. Since the analysis includes only Johnson-Nyquist noise, additional decoherence mechanisms present in a realistic experiment are expected to reduce this value.

Within our model, the coherence time is limited primarily by the gradual increase of the electron motional quantum number, occurring at a rate of approximately $\Gamma \approx \qty{30}{\per\second}$, as discussed in Sec.~\ref{sec:hrate}. One possible route toward extending the coherence time is to reduce the trap temperature and thereby suppress the electric-field noise. Another possibility is to periodically reinitialize the motional state through controlled recombination with an ion located at the trap centre. The investigation of such protocols and their application to quantum information processing represents an interesting direction for future work.

In conclusion, we have demonstrated a GHz microwave-resonator Paul trap fabricated by laser powder bed fusion that combines resonant enhancement of the trapping field with the large optical access required for hybrid electron-ion experiments. The measured resonator characteristics show good agreement with electromagnetic simulations despite the manufacturing imperfections inherent to additive fabrication. The validated model predicts trap depths sufficient for electron confinement and indicates that millisecond-scale electron-spin coherence should be achievable under the assumptions considered in this work. These results establish additive manufacturing as a viable route toward the realization of complex microwave traps for future charged-particle and quantum-technology applications. The method of obtaining the coherence time is, however, not limited to the case of additively manufactured traps and can be adapted to any electron Paul trap.

\begin{center} \textbf{AUTHOR CONTRIBUTIONS} 
\end{center}

\textbf{Niklas V Lausti}: Software (lead), Conceptualization (lead), Formal analysis (equal), Investigation (equal), Validation (lead), Visualization (supporting), Data curation (supporting), Writing – original draft (equal); \textbf{Vineet Kumar}: Conceptualization (equal), Data curation (lead), Formal analysis (lead), Investigation (lead), Visualization (lead), Validation (equal), Writing – original draft (lead); \textbf{Ji\v{r}\'{i} Hajny\v{s}}: Resources (equal), Writing – review \& editing (supporting); \textbf{Peter K\'{u}\v{s}}: Resources (equal), Writing – review \& editing (supporting); \textbf{Ivan Hud\'{a}k}: Data curation (supporting), Formal analysis (supporting); \textbf{Radek Pla\v{s}il}: Resources (equal), Writing – review \& editing (supporting); \textbf{Michal Hejduk}: Conceptualization (equal), Funding acquisition (lead), Project administration (lead), Resources (equal), Supervision (lead), Writing – review \& editing (lead)

\begin{center} \textbf{FUNDING} 
\end{center}

This work is supported by the Czech Science Foundation (GA\v CR: GA24-10992S), the Charles University Grant Agency (GAUK 295023 and GAUK 131224), and the Czech Ministry of Education, Youth, and Sports (project QM4ST, id. no. EH22\_008/0004572). Additional funding was provided by the Cooperatio Programme of Charles University. IH thanks to the
Technology Agency of the Czech Republic (TAČR: TN02000020) for the support. Furthermore, VK acknowledges the European Cooperation in Science and Technology (COST Action: CA17113, Ref.: E-COST-GRANT-CA17113-4b3f9042) for a Short-Term Scientific Mission grant, which facilitated training on a cryogenic linear Paul ion trap at the University of Liverpool.

\begin{center} \textbf{DATA AVAILABILITY} 
\end{center}

The data that support the findings of this article are publicly available\cite{laustiSupportingDataArticle2026}.

\begin{center} \textbf{ORCID IDs} \end{center}

Niklas Vilhelm Lausti

\url{https://orcid.org/0000-0001-9906-6971}

Vineet Kumar

\url{https://orcid.org/0000-0001-8668-3663}

Ivan Hud\'{a}k

\url{https://orcid.org/0009-0004-4162-6154}

Ji\v{r}\'{i} Hajny\v{s}

\url{https://orcid.org/0000-0002-9228-2521}

Peter K\'{u}\v{s}

\url{https://orcid.org/0000-0002-2246-4426}

Radek Pla\v{s}il

\url{https://orcid.org/0000-0001-8520-8983}

Michal Hejduk

\url{https://orcid.org/0000-0002-4417-4817}

\appendix

\section{Manufacturing and Dimensional Uncertainties}
\label{sec:dimensionerror}

The resonator was manufactured from AlSi10Mg aluminium alloy using a Renishaw AM500S Flex system equipped with a gas flow system. The printing process was conducted in an argon atmosphere (purity 5.0) to prevent oxidation. Recycled powder was reused up to four times during the process. The applied printing parameters were set according to Table~\ref{table:1}.

\begin{table}[h!]
\centering
\begin{tabular}{ccc}
\hline
\hline
\textbf{S.No.} & \textbf{PARAMETERS} & \textbf{VALUES} \\
\hline
1 & Laser power & 500 W \\
\hline
2 & Scan speed & 1875 mm s$^{-1}$ \\
\hline
3 & Layer thickness & \qty{30}{\micro\meter} \\
\hline
4 & Hatch distance & \qty{90}{\micro\meter} \\
\hline
5 & Laser spot size & \qty{75}{\micro\meter} \\
\hline
6 & Scanning strategy & Stripes \\
\hline
7 & Pre-heating & 170 $^\circ$C \\
\hline
\hline
\end{tabular}
\caption{3D-printing parameters for the resonator manufacturing.}
\label{table:1}
\end{table}

The aluminium powder used for L-PBF had a grain size of \qty{42\pm22}{\micro\meter}, which inherently produced a relatively coarse surface finish. Consequently, post-processing surface smoothing was necessary to achieve the surface quality required for optimal RF performance. 

Four resonator systems were fabricated to compare polishing methods: one unpolished reference and three treated by tumbling, electropolishing, and manual polishing. Manual polishing used FEPA P150 sandpaper and Dremel\textsuperscript{\textregistered} polishing wheels (EZ SpeedClic 472S, grit 120, and 511S, grit 180).

The diameters in the lower sections of part \textbf{J} (from Figure \ref{fig:1}) were measured by digital calliper. The unpolished specimen had a value \qty{5.01\pm0.02}{\milli\meter}, with no significant change after tumbling. Electropolishing and manual polishing reduced the diameters to \qty{4.95\pm0.02}{\milli\meter}, indicating material removal of \qty{60\pm40}{\micro\meter}. Therefore, the aggregated uncertainty of the dimension is estimated to be around \qty{100}{\micro\metre}.

\section{Surface Roughness Measurement}
\label{sec:surfaceroughness}

\begin{figure*}
    \centering
    \includegraphics{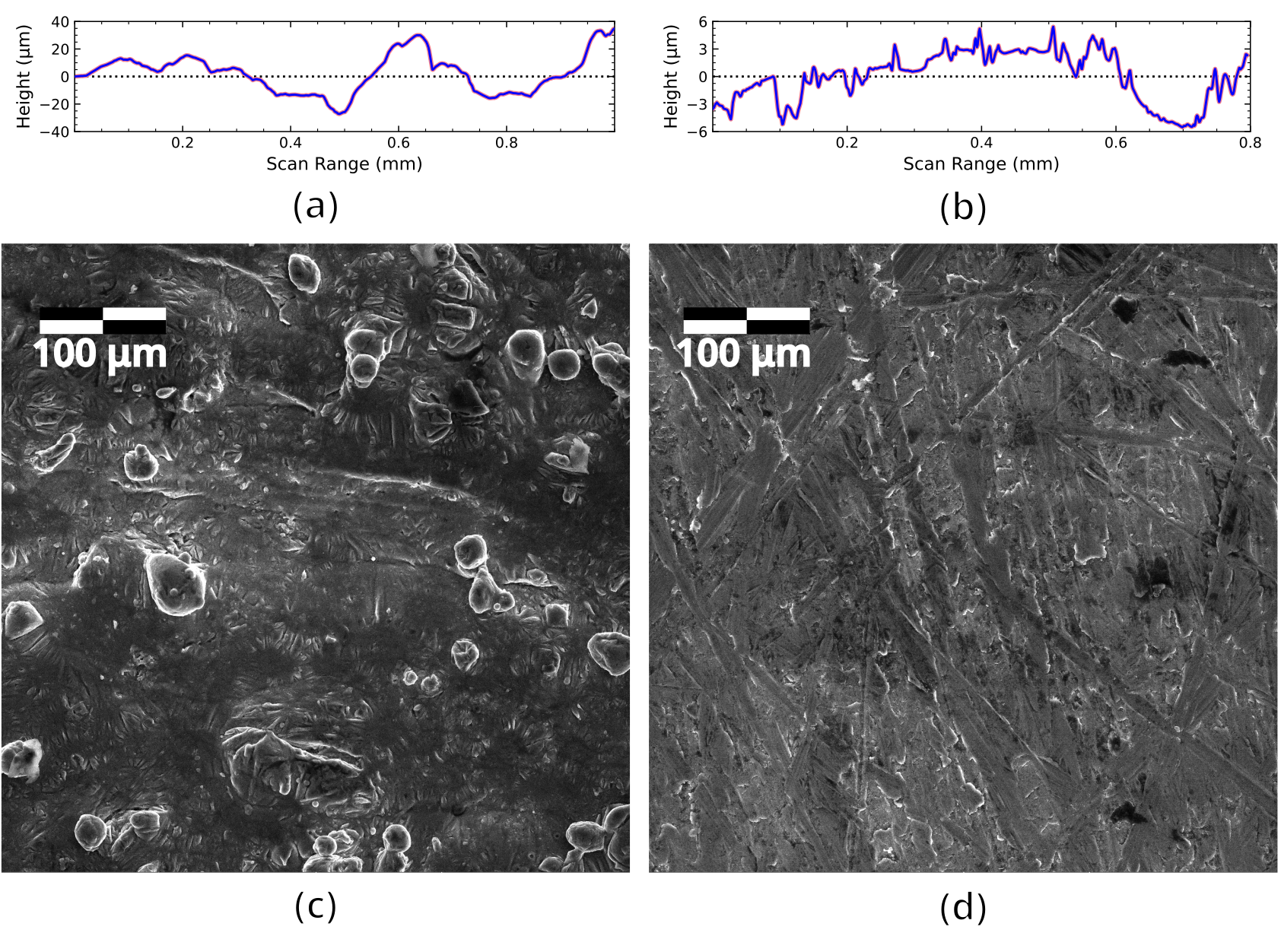}
\vspace{-2.0em}    
    \caption{Surface characterization of the trap sample. (a) and (b): Surface profilometry profiles over a 1 mm scan length of a sample before and after polishing, respectively. The measured rms surface roughness was $R_{qu}=$ \qty{14\pm10}{\micro\meter} for the unpolished surface and $R_{qp}=$ \qty{2.5\pm0.7}{\micro\metre} for the polished one. (c) and (d): SEM image of the samples before and after polishing, respectively. Large structures of the size of tens of \unit{\micro\metre} are effectively removed by polishing.} 
    \label{fig:SEM_SP}
\end{figure*}

Measurements with a contact stylus scanning profilometer (Dektak Pro) in Figure \ref{fig:SEM_SP}a showed that untreated 3D printed parts exhibited a root mean square roughness (rms) of \qty{14\pm 10}{\micro\meter}. These values are substantially higher than the electric skin depth of aluminium in the operating frequency range ($\approx$ \qty{2}{\micro\metre}).

The roughness after manual polishing was evaluated using an optical profilometer (Alicona InfiniteFocus G5). An example scan along a line is shown in Figure \ref{fig:SEM_SP}b. Calculating the average of five measured datasets, we found that the rms roughness of polished surfaces was \qty{2.5\pm0.7}{\micro\metre}. Some large-scale slopes on the surfaces have been excluded from the analysis, as they do not affect the microstructure of the surface.

The general morphology of the unpolished and polished surface was examined by scanning electron microscopy (SEM), and example images of them can be seen in Figures \ref{fig:SEM_SP}c,d. Although the surface of an unpolished sample was covered with globules of around \qty{30}{\micro\metre} size, such globules could not be found in the polished one.

\section{Material Composition Measurement}
\label{sec:composition}

\begin{figure}
    \centering
    \includegraphics[width=\linewidth]{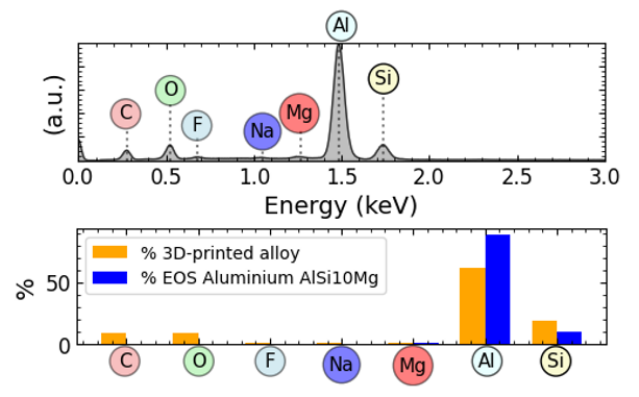}
    \caption{Energy-Dispersive X-ray spectrum of a sample printed out by Laser Powder Bed Fusion method (a) and comparison of material composition obtained from the spectrum with values tabulated for AlSi10Mg (b).}
    \label{fig:composition}
\end{figure}

The Energy-Dispersive X-ray Spectroscopy (EDS) was used to analyse the elemental composition of the material. As shown in Figure \ref{fig:composition}, we find that Al and Si contents  -- \(61.5\)~\% and \(18.7\)~\%, respectively -- deviate by \(27.5\)~\% and \(8.7\)~\% from the standard EOS AlSi10Mg composition. The lower Al content is likely influenced by surface effects and may be just a local phenomenon. Elevated C (9 \%) and O (8.5 \%) contents are attributed to surface contamination, while minor F and Na ($<1$ \%) may originate from handling or environmental sources. As the composition of the printout does not significantly deviate from the unprocessed alloy, we take the tabulated conductivity value of 25\% International Annealed Copper Standard (IACS), i.e.\ $\sigma = 1/\rho \approx \qty{1.45e7}{\siemens\per\metre}$.

 Due to the irregular geometry of the resonator, \(R_{\text{id}}\) from Equation \eqref{eq:resistance} is challenging to calculate accurately, but \(R_{\text{id}} \approx 25\text{ m}\Omega\) can be used as an estimate. (The resonator rod and a trap mount contribute approximately 21\ m$\Omega$, while the outer cavity can be assumed to have a minor contribution.) 

\section{Experimental Verification of Resonator}
\label{sec:verification}

 The resonator was characterized through direct power-spectrum measurements. For this purpose, the pickup probe (\(\textbf{H}_2\)) in figure~\ref{fig:1}b was positioned symmetrically opposite to the feeding antenna, rather than near the central trapping region (\(\textbf{K}\)), in order to weakly sample the standing-wave field while minimizing field perturbation and additional power extraction, avoiding artificial distortion of the resonance bandwidth, and preserving the electric-field maximum at the trap centre. In addition, the central region of the resonator is geometrically constrained by multiple apertures required for fluorescence light-cone collection from trapped \(^{40}\mathrm{Ca}^{+}/^{40}\mathrm{Ca}^{2+}\) ions, injection of optical-transition laser beams, and passage of the atomic beam source. The quality factor was extracted directly from the Lorentzian power spectrum using
\(Q=\frac{f_{0}}{\Delta f}\)
where \(f_{0}\) and \(\Delta f\) are the resonance frequency and FWHM, respectively. 

The quality factor was measured to be \num{990\pm10} at a resonance frequency of \(\Omega_\text{e}^{'} = 2.31\, \text{GHz}\). This value is below the simulated \num{1076\pm9}, likely because of signal reflections at the vacuum feedthrough interfaces and additional cavity-loading effects. These values correspond to a voltage amplification at the trapping region of approximately \(240\,\mathrm{V}\) for \qty{7.5\pm0.3}{V} input supplied to the feeding antenna \(\textbf{H}_1\), which is sufficient for electron trapping, as confirmed by FEM simulations.

As the present device is also intended for dual-frequency trap operation, FEM simulations were performed to estimate the electric field generated by the endcap electrodes. The simulated field strengths were found to be on the order of a few kilovolts per meter, in good agreement with the experimentally observed value of approximately \( 3\,\mathrm{kV/m}\). Since the primary focus of the present work is electron trapping, a more detailed characterization of the device from this perspective is provided in the work by \citet{kumar2026design}.

\section{Relation Between Trap Anharmonicity and Kerr Coefficient} \label{app:anharmonicity} 

For the derivation of the Kerr parameter, we adapt the method described by \citet{fengQuantumDissipativeDynamics2025} to our scenario.

The ponderomotive trapping potential \eqref{eq:elpot} along the z-axis is expressed as 
\begin{equation}
\Psi(z) = \frac{\me\omega_z^2}{2}z^2 + \beta z^4 + \mathcal{O}(z^6), \end{equation}
where $\beta$ is determined by the electrode geometry and quantifies the leading deviation from harmonic confinement. Introducing the ladder operators 
\begin{equation}
z = z_0(a_z+a_z^\dagger), \qquad z_0 = \sqrt{\frac{\hbar}{2\me\omega_z}}, 
\end{equation} 
the motional Hamiltonian becomes
\begin{equation} 
H_\mathrm{m} = \hbar\omega_z a_z^\dagger a_z + \beta z_0^4 (a_z+a_z^\dagger)^4. \end{equation} 

Expanding the quartic term and retaining only number-conserving contributions within the secular oscillation approximation yields 
\begin{equation}
(a_z+a_z^\dagger)^4 \approx 6(a_z^\dagger a_z)^2 + 12a_z^\dagger a_z + 3.
\end{equation} 

The last two terms merely renormalize the trap frequency and the overall zero-point energy. The leading nonlinear contribution is therefore 
\begin{equation} 
H_\mathrm{nl} = K_z(a_z^\dagger a_z)^2, 
\end{equation}
with the Kerr coefficient
\begin{equation}
K_z = 6\beta_z z_0^4 = \frac{3}{2} \frac{\beta_z\hbar^2} {\me^2\omega_z^2}. \end{equation} 

Consequently, the quartic coefficient $\beta_z$ directly determines the occupation-dependent frequency shift of the motional mode. Increasing $\beta_z$ increases the separation between neighbouring Fock-state transition frequencies.

\section{Derivation of Coherence Time}
\label{sec:coherencederiv}

Starting from
\begin{equation}
\frac{d c_n}{dt}
=\Gamma(n+1)c_{n+1}
+\Gamma n\,c_{n-1}
-\Gamma'n\,c_n
-\Gamma''c_n
\end{equation}
where
\begin{equation}
\Gamma'=2\Gamma+i\omega_z\eta_2,\, \Gamma'' = 
\Gamma(1-\eta_1^2)+i\omega_z\left(\eta_0
+\frac{\eta_1^2 \eta_2}{4}\right)
\end{equation}
we introduce the generating function
\begin{equation}
\mathcal{G}(z,t)=\sum_{n=0}^{\infty} c_n(t)z^n .
\end{equation}

Using
\begin{eqnarray}
\sum_{n=0}^{\infty}(n+1)c_{n+1}z^n
&=&
\partial_z\mathcal{G},
\\
\sum_{n=0}^{\infty}n\,c_{n-1}z^n
&=&
z(\mathcal{G}+z\partial_z\mathcal{G}),
\\
\sum_{n=0}^{\infty}n\,c_n z^n
&=&
z\partial_z\mathcal{G},
\end{eqnarray}
the recurrence relation is transformed into the first-order partial differential equation
\begin{equation}
\partial_t\mathcal{G}=a(z)\partial_z\mathcal{G}+b(z)\mathcal{G},
\end{equation}
where
\begin{eqnarray}
a(z) &=& \Gamma (z^2 - 2\beta z + 1) ,\\[1ex]
b(z)&=& \Gamma (z- \gamma) 
\end{eqnarray}
with
\begin{equation}
    \beta = \frac{\Gamma'}{2 \Gamma},\qquad\gamma = \frac{\Gamma''}{\Gamma}\,.
\end{equation}

Hence
\begin{equation}
\partial_t\mathcal{G} -  (\Gamma z^2 - \Gamma'z + \Gamma)\partial_z \mathcal{G} = (\Gamma z- \Gamma'')\mathcal{G}.
\end{equation}

The method of characteristics gives
\begin{equation}
\frac{dz}{dt}
=-\Gamma (z^2 - 2\beta z + 1).
\end{equation}

The quadratic polynomial factorizes as
\begin{equation}
 \Gamma(z^2 - 2\beta z + 1) = \Gamma(z-r_+)(z-r_-),
\end{equation}
with roots
\begin{equation}
r_\pm = \beta\pm\frac{\Delta}{2}\,,
\end{equation}
where $\Delta = 2\sqrt{\beta^2-1}$.

The characteristic equation therefore becomes
\begin{equation}
\label{eq:chareq}
\frac{dz}{dt} = -\Gamma(z-r_+)(z-r_-).
\end{equation}

Integrating this,
\begin{equation}
\frac{1}{r_+-r_-}
\ln\!\left(\frac{z-r_+}{z-r_-}\right) = -\Gamma t+C,
\end{equation}
where $C$ is constant along each characteristic. So the characteristic invariant may be written as
\begin{equation}
    \xi = [(z-r_+)/(z-r_-)]e^{\Delta\Gamma  t}\,.
\end{equation}

Along a characteristic,
\begin{equation}
\frac{d\mathcal{G}}{dt} = \Gamma(z- \gamma)\mathcal{G}.
\end{equation}

Using \eqref{eq:chareq}, we get
\begin{equation}
d\ln \mathcal{G} = -\frac{z- \gamma}{(z-r_+)(z-r_-)}\,dz = \frac{P}{z-r_+}\,dz + \frac{Q}{z-r_-}\,dz.
\end{equation}
where
\begin{equation}
P = -\frac{r_+-\gamma}{\Delta},
\qquad Q = \frac{r_- - \gamma}{\Delta}.
\end{equation}

Integrating give the general solution
\begin{equation}
\mathcal{G}(z,t) = (z-r_+)^P(z-r_-)^Q F\!\left(\frac{z-r_+}{z-r_-}e^{\Delta \Gamma t}\right),
\end{equation}
where \(F\) is determined by the initial condition.

For an initial Fock state, we suppose $c_n(0)=\delta_{n,n_0}$. Then
\begin{equation}
\mathcal{G}(z,0)=z^{n_0}.
\end{equation}

So, at $t=0$, we must have
\begin{equation}
    z^{n_0}=(z-r_+)^P(z-r_-)^Q F\!\left(\frac{z-r_+}{z-r_-}\right)
\end{equation}
and $\xi = (z- r_+)/(z-r_-)$. Solving the latter for $z$ gives
\begin{equation}
    z= \frac{r_+ - \xi r_-}{1-\xi}\,\text{ and } z-r_- = \frac{r_+ - r_-}{1-\xi}\,.
\end{equation}

Substituting into the initial condition and simplifying using the identity $P+Q = -1$ yields
\begin{equation}
F(\xi)= \Delta\,\xi^{-P}(1-\xi)^{-1}\left(\frac{r_+ - \xi r_-}{1-\xi}\right)^{n_0}     
\end{equation}

Now, to compute $C(t)=\mathcal{G}(1,t)$, and define the characteristic invariant at $z=1$:
\begin{equation}
    \xi_1(t)=\frac{1-r_+}{1-r_-}e^{\Delta\Gamma t}=-\left(\beta +\frac{\Delta}{2}\right)e^{\Delta\Gamma t}.
\end{equation}

Substituting this to the general solution yields
\begin{equation}
    C(t)=(1-r_+)^P (1-r_-)^Q F(\xi_1(t)).
\end{equation}

Using the expression for $F$ obtained above, we reach the result
\begin{equation}
\begin{split}
    \ln C(t)=
(r_+-\gamma)\Gamma t
+n_0\ln\!\left(\frac{r_+ + e^{\Delta\Gamma t}}{r_+ + 1}\right)\\
-(n_0+1)\ln\!\left(\frac{1+r_+e^{\Delta\Gamma t}}{1+r_+}\right)\,,
\end{split}
\end{equation}
which is rewritten using the original variables as in \eqref{eq:coherence}.

\bibliography{unified}% Produces the bibliography via BibTeX.
\end{document}